\documentstyle[prb,aps,epsf,eqsecnum]{revtex}

\begin{document}
\draft
\twocolumn[\hsize\textwidth\columnwidth\hsize\csname  
@twocolumnfalse\endcsname
\author{N. Andrenacci, P. Pieri, and G.C. Strinati}
\address{Dipartimento di Matematica e Fisica, Sezione INFM 
 Universit\`{a} di Camerino, I-62032 Camerino, Italy}
\date{\today}
\title{{\bf Size shrinking of composite bosons for increasing density in 
the BCS to Bose-Einstein crossover}}
\maketitle
\begin{abstract}
We consider a system of fermions in the continuum case at zero temperature, 
in the strong-coupling limit of a short-range attraction when composite bosons 
form as bound-fermion pairs.
We examine the density dependence of the size of the composite bosons at 
leading order in the density (``dilute limit''), and show on general physical 
grounds that this size should \emph{decrease\/} with increasing density, both 
in three and two dimensions.
We then compare with the analytic zero-temperature mean-field solution, which
indeed exhibits the size shrinking of the composite bosons both in three 
and two dimensions.
We argue, nonetheless, that the two-dimensional mean-field solution is not
consistent with our general result in the ``dilute limit'', to the extent that 
mean field treats the scattering between composite bosons in the Born 
approximation which is known to break down at low energy in two dimensions.
\end{abstract}
]

\section{Introduction}
The BCS to Bose-Einstein (BE) crossover can be regarded as an evolution from 
large overlapping Cooper pairs (BCS limit) to small nonoverlapping (composite) 
bosons (BE limit).
At zero temperature, this crossover has thus been characterized in the continuum 
case in terms of the correlation length $\xi_{{\rm pair}}$ for pairs of 
opposite-spin fermions (in units of $k_{F}^{-1}$, where $k_{F}$ is the Fermi 
wave vector).\cite{PS-94}

It was found that, by reducing the strength of the fermionic attraction at fixed 
density, $\xi_{{\rm pair}}$ increases monotonically from its strong-coupling (BE) limit 
(equal to the bound-state radius of the associated two-body problem) to the 
Pippard value in the weak-coupling (BCS) limit.

It was emphasized recently that, as the particle density increases for fixed 
value of the potential strength, the system also evolves from the BE limit toward 
the crossover region, eventually reaching the BCS limit (provided certain conditions 
on the fermionic attraction are fulfilled).\cite{APPS}

In this context, one would naively expect $\xi_{{\rm pair}}$ to increase monotonically, 
too, when evolving from the BE limit toward the crossover region by increasing 
the particle density.
A more careful analysis, however, shows that $\xi_{{\rm pair}}$ should actually 
\emph{decrease with increasing density\/} starting from the BE limit  
for small density (``dilute limit'').
Consideration of this effect, which originates from the repulsive interaction 
between the composite bosons, is the main purpose of this paper.

The composite bosons (which form as bound pairs from the constituent fermions in
the strong-coupling limit) mutually interact via a residual repulsive interaction
due to Pauli principle.\cite{Haussmann,PS-96}
This repulsive interaction, in turn, affects the size of the composite bosons.
Specifically, one expects on physical grounds this interaction to \emph{reduce\/} 
the size of the composite bosons (with respect to the bound-state radius of a 
composite boson in isolation), insofar as the repulsive interaction itself
decreases for decreasing size of the composite bosons.
A new equilibrium size for the composite bosons thus results when combining the 
effect of the repulsive interaction with the internal energy of a composite boson, 
the new equilibrium size being smaller than the original size of a composite boson 
in isolation.

We shall implement this qualitative idea by minimizing the expression of (twice) 
the fermionic chemical potential $2 \mu \, = \, - \epsilon \, + \, \mu_{B}$ with 
respect of the size $\xi_{{\rm pair}}$ of a composite boson, where $\epsilon$ represents
the \emph{internal\/} energy of a composite boson (to be defined below) and 
$\mu_{B}$ is the bosonic chemical potential determined by the mutual repulsion 
between the composite bosons.\cite{KK}
Both $\epsilon$ and $\mu_{B}$ are, in fact, functions of $\xi_{{\rm pair}}$; 
in addition, $\mu_{B}$ depends on the density $n$.

We shall explicitly verify that the size-shrinking effect is borne out by 
the analytic zero-temperature mean-field solution for a point-contact 
interaction both in three\cite{MPS} and two\cite{Randeria-90} dimensions.
In two dimensions, however, the mean-field density dependence of $\xi_{{\rm pair}}$
does not agree with what expected in the ``dilute limit''.
We attribute this difference to the poor treatment of the bosonic scattering
within the zero-temperature mean field, which rests on the Born approximation
as the form of the mean-field bosonic chemical potential implies.
Since the Born approximation in two dimensions is known to fail at low energy
(which, in turn, corresponds to the ``dilute limit'' of the Bose gas), the effect 
of the bosonic interaction on $\xi_{{\rm pair}}$ cannot be properly treated within the 
Born approximation.

We will therefore argue that a proper treatment of the residual bosonic interaction 
(over and above the Born approximation), along the lines recently developed for the 
three-dimensional case,\cite{Pi-S-98} should be especially relevant to the 
two-dimensional case.

We have focused in this paper on the composite-boson regime of the BCS-BE
crossover problem, following previous work on this problem 
\cite{PS-96,Pi-S-98} which has emphasized the importance of approaching the
bosonization {\em in reverse},
that is, starting from the BE and evolving toward the BCS region. 
Although this 
reverse approach is not usually adopted in the literature, we prefer it over
the conventional BCS to BE approach which starts from the BCS limit, insofar as
the physics of the BE (composite-boson) region is much richer than its 
counterpart in the BCS limit. Specifically, the reverse approach enables one to
perform approximations in the BE limit which would be {\em a priori} 
uncontrolled if one would start instead from the BCS limit. These 
approximations, which are necessarily more involved than those performed in the
BCS limit, are nevertheless appropriate also to the BCS limit.\cite{Pi-S-98}
In addition, it has been shown that even numerical approximation schemes 
starting from the BE region are surprisingly accurate also on the BCS side of
the crossover, while the reverse is not true.\cite{MPS}

The plan of the paper is as follows.
In Section 2 we provide the general physical argument for the size shrinking 
of the composite bosons with increasing density in the ``dilute limit'', 
irrespective of dimensionality.
In Section 3 we confront this general argument with the analytic 
zero-temperature mean-field results for a contact potential in the 
strong-coupling limit, both in three and two dimensions. Numerical results for
two different finite-range potentials in three dimensions are also shown for 
comparison.
Section 4 gives our conclusions.
\section{General argument for the size shrinking of the composite bosons}

We consider a system of fermions at zero temperature, mutually interacting via 
an (effective) attractive potential with a finite-range $r_0$.
For the sake of comparison with the available analytic results in three and two
dimensions, we disregard lattice effects and consider the system embedded
in a homogeneous background (continuum case).

Composite bosons in isolation are defined when the associated two-body 
problem admits a bound state (zero-density limit).
At small (albeit finite) density, composite bosons are expected to retain
their identity in the strong-coupling limit of the original fermionic
attraction, i.e., when the binding energy of the associated two-body problem
is much larger than the strength of the \emph{residual interaction\/}
between the composite bosons.
Quite generally, this residual interaction has a dominant short-range 
\emph{repulsive\/} part due to Pauli principle, which is active among the
constituent fermions as soon as the composite bosons overlap.
This overlap, in turn, increases (on the average) when the size and/or the
density of the composite bosons increase.

More precisely, we shall assume that $k_F a_F << 1$, where $a_F$ is the 
scattering length associated with the fermionic attraction, as well as
$a_F >> r_0$.
The former assumption represents a ``diluteness'' condition, while the second 
assumption is required to get a well-defined system of composite bosons, for
which the Pauli repulsion overwhelms the attractive part of the bosonic 
potential originating from the finite-range fermionic attraction (see 
Ref.~[8], footnote [36]). In this way, instabilities of the bosonic system 
(like the one recently pointed out in Ref.~[9]) will be suitably 
avoided.

When the above conditions are satisfied, the composite bosons have
a finite size and yet can be considered to be well-defined entities.
We assume, therefore, that the standard results for the Bose gas can be used 
as far as the mutual interaction of the composite bosons is concerned, 
irrespective of their internal structure.
To the internal structure we associate a pair wave function $\psi$ and a 
corresponding internal energy $\epsilon$, obtained from (minus) the expectation 
value $\langle \psi | H_{2} | \psi \rangle / \langle \psi | \psi \rangle$ of the 
two-fermion Hamiltonian $H_{2}$ over the pair wave function (where $H_{2}$ contains 
the reduced kinetic energy and the attractive two-fermion potential).
Physically, $- \epsilon$ represents the energy required to form a composite
boson in isolation \emph{with a given\/} internal wave function $\psi$ for the
pair of constituent fermions.
We thus regard the energy $2 \mu$ required to add \emph{two\/} fermions to 
the system as composed of two distinct contributions, namely, \emph{(i)}
the (negative of the) above internal energy $- \epsilon$ and \emph{(ii)}
the energy $\mu_{B}$ required to add eventually the composite boson to the 
system (with a \emph{finite density\/} of composite bosons already present).
On physical grounds, $\mu_{B}$ is related to the repulsive interaction between 
the composite bosons and depends on their size and density.
Specifically, in three dimensions we take (we set $\hbar=1$ throughout)

\begin{equation}
\mu_{B} \, = \, \frac{4 \pi n_{B} a_{B}}{m_{B}}                    
\label{mu-B-3}
\end{equation}

\noindent
from the standard theory of the ``dilute'' Bose gas,\cite{FW} where $n_{B}=n/2$ 
is the bosonic density, $m_{B}=2m$ is the mass of a composite boson in terms 
of the mass $m$ of the constituent fermions, and $a_{B}$ is the (positive) 
bosonic scattering length due to the repulsive interaction between the composite 
bosons.
In two dimensions we take instead

\begin{equation}
\mu_{B} \, = \, \frac{4 \pi n_{B}}{m_{B} \ln\left(\frac{1}{n_{B} r_{o}^{2}}\right)}
                                                             \label{mu-B-2}
\end{equation}

\noindent
from the theory of the two-dimensional ``dilute'' Bose gas,\cite{FH} where
$r_{o}$ represents the range of the bosonic interaction.
From dimensional considerations, both $a_{B}$ in Eq.~(\ref{mu-B-3}) and $r_{o}$ 
in Eq.~(\ref{mu-B-2}) should be proportional to the size $\xi_{{\rm pair}}$ of the 
composite bosons, defined by the following functional:

\begin{equation}
\xi_{{\rm pair}}^{2}[\psi] \, = \, 
\frac{ \int \! d {\mathbf r} \,\, |\psi({\mathbf r})|^{2} \, {\mathbf r}^{2} }
     { \int \! d {\mathbf r} \,\, |\psi({\mathbf r})|^{2} } \,\, .     
                                                             \label{xi-pair}
\end{equation}

\noindent
We set correspondingly $a_{B} = \alpha_{3} \, \xi_{{\rm pair}}$ and 
$r_{o} = \alpha_{2} \, \xi_{{\rm pair}}$ in three and two dimensions, in the order,
where $\alpha_{3}$ and $\alpha_{2}$ are positive constants.
In this way, the bosonic chemical potential $\mu_{B} = \mu_{B}(\xi_{{\rm pair}}$) 
becomes itself a function of $\xi_{{\rm pair}}$.

We determine next the function $- \epsilon(\xi_{{\rm pair}})$ by minimizing
$\langle \psi | H_{2} | \psi \rangle / \langle \psi | \psi \rangle$ in the
subspace of the wave functions $\psi$ corresponding to a given value of
$\xi_{{\rm pair}}$.
In this way, the equilibrium value $\xi_{{\rm pair}}^{(o)}$ and the associated internal 
energy $\epsilon_{o}$ are obtained from the global minimum of $- \epsilon(\xi_{{\rm pair}})$
($\epsilon_{o}$ coinciding with the binding energy of the associated two-body
problem).
Near this minimum we set

\begin{equation}
\epsilon(\xi_{{\rm pair}}) \, = \, \epsilon_{o} \, - \, A_{d} \, 
\left( \xi_{{\rm pair}} \, - \, \xi_{{\rm pair}}^{(o)} \right)^{2}         \label{eps-xi}
\end{equation}

\noindent
where the positive constant $A_{d}$ depends in general on the dimensionality $d$.
This approximate expression can be used when searching for the minimum of

\begin{equation}
2 \, \mu(\xi_{{\rm pair}}) \, = \, - \, \epsilon(\xi_{{\rm pair}}) 
                     \, + \, \mu_{B}(\xi_{{\rm pair}})               \label{2mu(xi)}
\end{equation}
 
\noindent
to determine the new equilibrium value $\bar{\xi}_{pair}$ in the presence of a
finite (albeit \emph{small\/}) density of composite bosons.
From Eq.~(\ref{mu-B-3}) we then obtain for this new equilibrium value in three 
dimensions:

\begin{equation}
\bar{\xi}_{pair} \, = \, \xi_{{\rm pair}}^{(o)} \, - \, \frac{2 \pi \alpha_{3}}{m_{B} A_{3}}
                 \,\, n_{B} \,\, ;                                 \label{xi-3}
\end{equation}

\noindent
while from Eq.~(\ref{mu-B-2}) we obtain in two dimensions:

\begin{equation}
\bar{\xi}_{pair} \, = \, \xi_{{\rm pair}}^{(o)} \, - \, 
\frac{\pi}{m_{B} \, A_{2} \, \xi_{{\rm pair}}^{(o)} 
\left[\ln\left(\alpha_{2} \, \xi_{{\rm pair}}^{(o)} \, n_{B}^{1/2}\right)\right]^{2}}
                 \,\, n_{B} \,\, .                                 \label{xi-2}
\end{equation}

\noindent
Note that the value of $\bar{\xi}_{pair}$ in the presence of a finite density 
of composite bosons is \emph{smaller\/} than the value $\xi_{{\rm pair}}^{(o)}$ for 
a composite boson in isolation, \emph{both\/} in three and two dimensions.
Physically, this shrinking is due to the fact that the decrease of the repulsive 
bosonic interaction (when the composite bosons contract) prevails over
the reduction of the internal energy away from the original equilibrium value 
$\epsilon_{o}$.

In three dimensions, by inserting Eq.~(\ref{xi-3}) into Eq.~(\ref{eps-xi}), 
the difference $\epsilon(\bar{\xi}_{pair}) - \epsilon_{o}$ is seen to decrease 
quadratically with increasing density.
The leading-order correction to the internal energy $\epsilon$ is then 
quadratic in the density. This should be contrasted with the density dependence
of the bosonic chemical potential, which is instead linear to the leading 
order.
It can be further checked that adding to equation (1) for the bosonic chemical
potential terms of higher order in the small parameter $n_B^{1/3} a_B$, 
will not modify expression (\ref{xi-3}) for the leading-order dependence of
$\xi_{pair}$ on $n_B$ in the dilute limit. Both the bosonic chemical potential
$\mu_B$ and the internal energy $\epsilon$ are thus self-consistently 
calculated by our approach to the leading order in the density.

To leading order, the \emph{binding energy\/} $- 2 \mu(\bar{\xi}_{pair})$
for a composite boson \emph{embedded in the medium\/} thus \emph{decreases 
linearly with the density\/}, owing to the linear dependence of the bosonic 
chemical potential.
This effect has been evidenced in different contexts, namely, for the Bose 
condensation of excitons in semiconductors \cite{KK} and, more 
recently,\cite{Kleppner} for the Bose condensation of atomic hydrogen.
In the latter case, a linear reduction of the bosonic binding energy has been
measured for increasing density, consistently with the general argument 
presented in this paper.

\section{Comparison with 3-d and 2-d mean-field results}
In this Section, we consider without loss of generality an attractive fermionic
point-contact potential, for which the analytic solution of the BCS to BE 
crossover at the zero-temperature mean-field level is available, both in three 
\cite{MPS} and two \cite{Randeria-90} dimensions.
We verify that this solution yields a decrease of $\xi_{{\rm pair}}$ for 
increasing density, both in three and two dimensions, in generic agreement 
with the results of the previous Section.
Specifically, in three dimensions we are able to recover the mean-field density
dependence of $\xi_{{\rm pair}}$ from the approach of the previous Section, by 
relating $a_{B}$ to $\xi_{{\rm pair}}$ via the Born approximation.
The analogous attempt fails, however, in two dimensions because the Born 
approximation (which is associated with mean field) strongly overestimates the
scattering between composite bosons, thus disrupting the basic assumptions on 
which the approach of the previous Section rests.

We remark that the analytic results presented in this Section for a 
point-contact interaction describe also the general behaviour for a 
{\em finite-range} potential in the ``dilute'' composite-boson regime, which is
defined by the two conditions $k_F a_F << 1$ and $a_F >> r_0$ previously 
mentioned. In this case, the range $r_0$ of the potential is much smaller than 
the two other length scales in the problem ($k_F^{-1}$ and $a_F$), so that the
mean-field equations for a finite-range potential get always mapped onto the 
mean-field equations for a point-contact interaction (with the same scattering
length). The region where the size-shrinking effect is present (namely, the 
composite-boson region) coincides thus with the region where the 
behaviour of the point-contact interaction is ``universal''. It is just this 
coincidence which enables us to establish the size-shrinking effect as a
general feature of the strong-coupling limit of the BCS-BE crossover.

\subsection{Three-dimensional case}

For a three dimensional point-contact potential, the low-energy fermionic 
two-body scattering can be conveniently regularized in terms of the scattering
length $a_{F}$.
At finite fermionic density, it is then possible to express all relevant 
physical quantities (such as the superconducting gap $\Delta$ and the chemical
potential) 
in terms of the dimensionless parameter $k_{F} a_{F}$.
This can be explicitly verified for the analytic solution of Ref.~[6] 
given in terms of the complete elliptic integrals.
Since $k_{F} a_{F} << 1$ in the BE limit, physical quantities can further be 
expanded in powers of $k_{F} a_{F}$ in this limit.

In particular, for the pair coherence length one finds from the analytic 
solution of Ref.~[6]:

\begin{equation}
\xi_{{\rm pair}} \, = \, \frac{a_{F}}{\sqrt{2}} \, \left[ 1 \, - \, \frac{5}{6\pi} 
                \, (k_{F} a_{F})^{3} \right]                      \label{MF-xi}
\end{equation}

\noindent
at the leading order in the bosonic density $n_{B} = k_{F}^{3}/(6 \pi^{2})$.
Equation [8] has been obtained by first expanding equation (30) of Ref.~[6]
in powers of $1/x_0$, by then inverting it to abtain $x_0$ in powers of 
$k_F a_F$, and by inserting the resulting expression for $x_0$ into the power
expansion of equation (31) of Ref.~[6] for $k_F \xi_{pair}$.
In equation (8) (as well as in equation (31) of Ref.~[6]) $\xi_{{\rm pair}}$ 
is defined according to equation (\ref{xi-pair}), with the BCS choice 
$\psi_{{\rm BCS}}({\mathbf k}) = 
\Delta/(2 E({\mathbf k}))$ for the Fourier transform (with wave vector 
${\mathbf k}$) of the pair wave function,\cite{PS-94} where $E({\mathbf k}) = 
\sqrt{({\mathbf k}^{2}/(2m) - \mu)^{2} + \Delta^{2}}$ as usual.

Note from Eq.~(\ref{MF-xi}) that in the zero-density limit $\xi_{{\rm pair}}$ equals the 
value $a_{F}/\sqrt{2}$ of the bound-state radius of a composite boson in isolation.
Note also that $\xi_{{\rm pair}}$ \emph{decreases linearly with increasing density\/} 
of the composite bosons. The mean-field analytic solution for a point-contact 
interaction thus confirms our general prediction of the size-shrinking effect
discussed in the previous Section [cf. Eq.~(\ref{xi-3})]. 

By the same token, for the bosonic chemical potential one finds from the 
analytic solution of Ref.~[6]:

\begin{equation}
\mu_{B} \, = \, 
\frac{2 \, a_{F} \, k_{F}^{3}}{3 \, \pi \, m} \, = \, 
\frac{8 \, \pi \, n_{B} \, a_{F}}{m_{B}} \, \, .                 \label{MF-B-3}
\end{equation}

\noindent
Comparison with Eq.(\ref{mu-B-3}) suggests then to identify $a_{B} = 2 a_{F}$
in the BE (strong-coupling) limit.
This result, which was also obtained for the same model system within the fermionic 
T-matrix approximation in the normal state (i.e., above the superconducting 
critical temperature),\cite{Haussmann,PS-96} identifies $k_{F}a_{F}$ with 
$(3 \pi^{2}/4)^{1/3}$ times the ``gas parameter'' $n_{B}^{1/3} a_{B}$ for a 
``dilute'' Bose gas.

Recall further that the general mapping procedure from the original fermionic
system onto the effective bosonic system in the strong-coupling limit 
(as described in Ref.~[4]) provides in three dimensions the value

\begin{equation}
v(0) \, = \, \frac{4 \, \pi \, a_{F}}{m}                     \label{v-3-0}
\end{equation}

\noindent
for the strength of the ``bare'' bosonic potential (with all wave vectors and 
Matsubara frequencies set to zero).
One then verifies that the Bogoliubov result

\begin{equation}
\mu_{B} \, = \, n_{B} \, v(0)                                \label{Bogoliubov}
\end{equation}

\noindent
is retrived by the expression (\ref{MF-B-3}).\cite{footnote-1} 

It is interesting to show that not only the generic linear density 
dependence but also the coefficients of the expression (\ref{MF-xi}) can be 
reproduced by the variational principle of the previous Section, \emph{provided\/}
we take for $\alpha_{3}$ the value $2 \sqrt{2}$ (as determined from 
Eqs.~(\ref{MF-B-3}), (\ref{MF-xi}), and (\ref{mu-B-3}) to leading order in the 
density) and we assume the BCS form 
$\psi_{{\rm BCS}}({\mathbf k}) = \Delta/(2 E({\mathbf k}))$ for the Fourier 
transform of the pair wave function.
To this end, it is convenient to express initially 
$\langle \psi | H_{2} | \psi \rangle / \langle \psi | \psi \rangle$ and $\mu_{B}$
as functions of the dimensionless variable $x_{o} = \mu/\Delta$ instead of
$\xi_{{\rm pair}}$, to minimize then the resulting expression for $2 \mu$ with respect to 
$x_{o}$, and to use eventually the functional relation between $x_{o}$ and $\xi_{{\rm pair}}$ 
(as determined in Ref.~[6]) to obtain $\xi_{{\rm pair}}$ in terms of $k_{F} a_{F}$.
In the BE limit, where $x_{o}<0$ and $|x_{o}|>>1$, this relation reads:

\begin{equation}
k_{F} \xi_{{\rm pair}} \, = \, \frac{1}{\sqrt{2}} \, 
      \left( \frac{3 \pi}{16 x_{o}^{2}} \right)^{1/3} \,
      \left( 1 \, - \, \frac{1}{4 x_{o}^{2}} \right)          \label{xi-x-o}
\end{equation}

\noindent
to the leading significant orders.
One finds:

\begin{eqnarray}
\frac{\langle \psi_{{\rm BCS}} | H_{2} | \psi_{{\rm BCS}} \rangle}
     {\langle \psi_{{\rm BCS}} | \psi_{{\rm BCS}} \rangle}
& = & \frac{k_{F}^{2}}{2m} \left( \frac{16}{3 \pi} \right)^{2/3} 
\left[ 2 \,|x_{o}|^{4/3} -  4\, b\, |x_{o}|^{2/3} \right. \nonumber \\
& + &\left. \frac{5}{8} \, |x_{o}|^{-2/3}
\, - \, \frac{5}{8} \, b \, |x_{o}|^{-4/3} \right]                   
\label{H-2-BCS}
\end{eqnarray}

\noindent
where we have set $b=(3\pi/16)^{1/3} \, (k_{F}a_{F})^{-1}$, while 
$\mu_{B}=\mu_{B}(x_{o})$ is given by Eq.~(\ref{mu-B-3}) with
$a_{B}=2\sqrt{2} \, \xi_{{\rm pair}}$ and with $\xi_{{\rm pair}}=\xi_{{\rm pair}}(x_{o})$ given 
by Eq.~(\ref{xi-x-o}).

At the leading order, only the first two terms within brackets in Eq.~(\ref{H-2-BCS})
are relevant to the expression of $2\mu$, whose mimimum is thus located at
$|\bar{x}_{o}| \, = \, b^{3/2}$.
At the next significant order, all terms within brackets in Eq.~(\ref{H-2-BCS}) and
the leading term in Eq.~(\ref{xi-x-o}) are relevant to the expression of $2\mu$,
yielding the new minimum at 

\begin{equation}
|\bar{x}_{o}| \, = \, b^{3/2} \, \left( 1 \, - \, \frac{9}{64 \, b^{3}} \right)
              \,\, .                                             \label{x-o-bar}
\end{equation}

\noindent
Inserting Eq.~(\ref{x-o-bar}) into Eq.~(\ref{xi-x-o}) one recovers eventually
the expression (\ref{MF-xi}) for $\xi_{{\rm pair}}$, as anticipated.
The fact that equation (8), obtained by the mean-field solution, can be 
reproduced {\em with the correct numerical coefficients} by the minimization 
procedure of the previous Section (when specialized to the BCS choice for the
pair wave function $\psi$), can be regarded as quite a compelling check on the
validity of our general argument of Section 2 and of its underlying 
assumptions. 

Finally, it is interesting to examine the behaviour of $\xi_{pair}$ as a 
function of the density for a finite-range potential, for which the BCS-BE 
crossover driven by the density becomes possible \cite{APPS}. As in Ref.~[2], 
we consider the separable Nozi\`eres-Schmitt-Rink (NSR) potential 
$V(k,k')=V(1+k^2/k_0^2)^{-1/2} (1+k'^2/k_0^2)^{-1/2}$ (with $k=|{\mathbf k}|$),
and the non-separable Gaussian potential 
$V({\mathbf k},{\mathbf k'})=V  \exp(-|{\mathbf k} - {\mathbf k'}|^2/k_0^2)$
($V<0$ in both cases). 
For these potentials, an analytic 
solution for the BCS-BE crossover is lacking, even at the mean-field level.
We have therefore solved numerically the coupled equations for the gap 
function $\Delta(k)$ and the chemical potential $\mu$. The pair-coherence 
length $\xi_{pair}$ has been then determined by using the Fourier transform of 
equation (3), with the BCS choice for the pair wave function 
$\psi(k)=\Delta(k)/(2 E(k))$. Results are shown in Figure 1, which confirm the
presence of the size-shrinking effect in the ``universal''  composite-boson 
region. Note, however, that the size-shrinking effect 
gradually disappears by increasing further the density: $\xi_{pair}$ reaches a
minimum at $k_F\sim k_0$ and then starts increasing with the density. The 
eventual increase of $\xi_{pair}$ with $k_F$, when $k_F>>k_0$, is a 
characteristic feature of a finite-range potential, as it can be readily 
verified from the asymptotic expressions (4) and (5) of Ref.~[2] for the gap
$\Delta(k_F)$ in the limit $k_F>>k_0$. In this limit, as soon as the system 
enters the BCS regime, we may use the BCS result $\xi_{pair}\sim 
k_F/\Delta(k_F)$ to obtain $\xi_{pair}\sim k_F^{-1} \exp\left(
\frac{k_F A}{| V | m k_0^2}\right)$, where $A$ is a numerical factor different
for the NSR and Gaussian potentials. For both potentials, $\xi_{pair}$ is seen 
to increase exponentially with $k_F$ for large $k_F$, as far as $k_0$ is 
finite.
The competition between this asymptotic behaviour of $\xi_{pair}$ when 
$k_F>>k_0$ and the initial size-shrinking when $k_F<<k_0$, leads then to 
a minimum of $\xi_{pair}$ when $k_F\sim k_0$.
\begin{figure}
\narrowtext
\epsfxsize=8truecm
\epsfbox{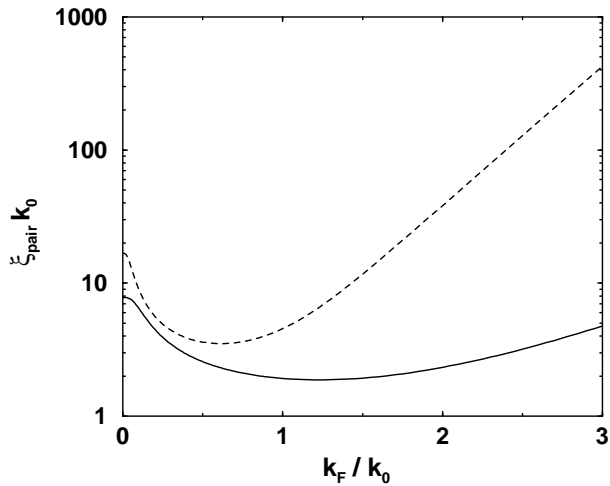}
\caption{Pair coherence length (in units of the characteristic length 
$k_0^{-1}$ of the finite-range fermionic potential) vs $k_F/k_0$, for the NSR 
(full line) and Gaussian (dashed line) potentials. [A typical value of 
$|V|$ ($=1.1 \;V_{c}$) above
the critical value $V_c$ for the existence of a bound state in three 
dimensions has been considered for both potentials.]}
\end{figure}

\subsection{Two-dimensional case}
In two dimensions, the zero-temperature mean-field expression for $\xi_{{\rm pair}}$ is 
reported in Appendix B of Ref.~[6] in terms of the available analytic 
solution.\cite{Randeria-90}
In particular, at the leading significant order in the BE limit one finds:

\begin{equation}
\xi_{{\rm pair}}^{2} \, = \, \frac{2}{3 m \epsilon_{o}} \, \left[ 1 - \frac{38}{15} \, 
\left(\frac{4 \pi n_{B}}{m_{B} \epsilon_{o}}\right) \right]  \,\, .     \label{MF-xi-2}
\end{equation}

\noindent
For given value of $\epsilon_{o}$, $\xi_{{\rm pair}}$ is thus seen to decrease linearly
with the density in the BE limit.
One also finds for the bosonic potential:

\begin{equation}
\mu_{B} \, = \, \frac{8 \, \pi \, n_{B}}{m_{B}}                                \label{MF-B-2}
\end{equation}

\noindent
which is proportional to the density but independent from the bosonic size.
Note from Eq.~(\ref{MF-B-2}) that the Bogoliubov result 
$\mu_{B} = n_{B} \, v(0)$ is retrieved by the mean-field calculation even in 
two dimensions, since the strength $v(0)$ of the ``bare'' bosonic potential is 
given by 

\begin{equation}
v(0) \, = \, \frac{8\pi}{m_{B}}  \,\, ,                        \label{v-2-0} 
\end{equation}

\noindent
as it can be explicitly verified by applying the prescriptions of 
Ref.~[4] to the fermion-boson mapping in the two-dimensional case.

The mean-field expressions (\ref{MF-xi-2}) and (\ref{MF-B-2}) differ from the
``dilute'' Bose gas expressions (\ref{xi-2}) and (\ref{mu-B-2}), respectively, as 
they lack the logarithmic term in the denominator.
Notwithstanding the decrease of the size of the composite bosons for increasing 
density obtained by Eq.~(\ref{MF-xi-2}), the two-dimensional mean-field results 
appear thus to contradict the picture of Section II for a ``dilute'' gas of 
composite bosons.
In particular, if one would use the Bogoliubov expression (\ref{MF-B-2}) to 
implement the argument of Section II, the failure of the strength of the 
``bare'' bosonic potential in two dimensions to depend on the size of the 
composite 
bosons would not make it energetically convenient to shrink the bosonic size 
at finite density.
The size shrinking obtained by Eq.~(\ref{MF-xi-2}) is therefore not consistent 
with the general argument developed in Section II.

We attribute the difference between the mean-field and the ``dilute'' gas results to 
the poor treatment of the boson-boson scattering within the zero-temperature mean 
field, which rests on the Born approximation as the form (\ref{MF-B-2}) of the 
bosonic chemical potential implies.
Let us, in fact, analyze the above results in terms of the outcomes of potential 
scattering theory in two dimensions.\cite{A}
The Born approximation gives for the (dimensionless) low-energy scattering amplitude
$f_{\mathrm{Born}}^{(2)} \sim - m v(0)$, where $v(0) \sim v_{o} r_{o}^{2}$ is typically 
proportional to the average strength $v_{o}$ and range $r_{o}$ of the potential.
The \emph{exact\/} low-energy result for the two-dimensional scattering amplitude
$f^{(2)} \sim -1/\ln(k r_{o})$, on the other hand, is independent of $v_{o}$ and
vanishes for vanishing wave vector ($k \rightarrow 0$).
In two dimensions, therefore, the effect of summing an infinite number of repeated
scatterings \emph{drastically modifies\/} the functional dependence of the scattering
amplitude on the strength and range of the potential, and the perturbation theory
for the scattering amplitude breaks down.\cite{LL}
In contrast, in three dimensions $f_{\mathrm{Born}}^{(3)} \sim - m v(0) \sim - 
a_{\mathrm{Born}}$ where $v(0) \sim v_{o} r_{o}^{3}$ and $a_{\mathrm{Born}}$ is the 
Born scattering length, whereas $f^{(3)} \sim - a$ is the \emph{exact\/} low-energy 
result ($a$ here being the full scattering length).
That is, in three dimensions going from the Born approximation to the exact 
low-energy result merely changes the numerical value of the scattering length.

For the composite bosons we are specifically interested in, the strength $v_{o}$
must be proportional to the binding energy $\epsilon_{o}$ for dimensional reasons,
and the only available length scale is proportional to $(m \epsilon_{o})^{-1/2}$.
This yields $v(0) \sim \epsilon_{o} (m \epsilon_{o})^{-d/2}$, which 
corresponds to Eqs.~(\ref{v-3-0}) and (\ref{v-2-0}) for $d=3$ and $d=2$, 
respectively, implying that the characteristic strength and range of the 
potential cannot be independently varied for the composite bosons.
It is this feature which, in turn, makes $v(0)$ independent from the size of 
the composite bosons in two dimensions.
The Born approximation to the two-dimensional scattering amplitude for the 
composite bosons, therefore, not only lacks the functional form of the exact 
low-energy result, but even fails to yield any dependence on the size of the 
composite bosons whatsoever.

From the above considerations, it is thus evident that in two dimensions
the scattering rate between the composite bosons is overestimated within the 
Born approximation with respect to the true low-density result, in such a way 
that the ``dilute limit'' cannot be achieved at fixed density.
Consequently, the physical picture adopted in Section II of a system of 
well-defined weakly interacting bosonic entities is bound to break down 
within mean field.

\section{Concluding Remarks}
In this paper, we have discussed the size shrinking of composite bosons as a 
general physical result occurring in the ``dilute limit''.
We have also verified that this general result is correctly reproduced by the 
zero-temperature mean-field treatment in three but not in two 
dimensions.
We have related the failure in two dimensions to the break down of the Born
approximation at low energy.
In addition, we have noted a peculiar relation between the effective strength
and range of the potential acting between composite bosons.

For the three-dimensional case, the relevance of including the mutual 
interaction between the composite bosons in the strong-coupling (BE) limit has
been emphasized some time ago in Refs.~[5] and [16]; more recently, 
the significance of treating the scattering between the composite bosons 
beyond the Born approximation has been addressed in Ref.~[8].
We have argued here, however, that the failure to account for the effects of 
the bosonic interaction is more severe in two than in three dimensions, because
the Born approximation breaks completely down in two dimensions.
By the same token, we also expect that the boundary in the phase diagram 
between 
the Bose-Einstein and crossover regions, discussed in Ref.~[2] within mean 
field, should be significantly modified by a proper treatment of the bosonic 
interaction, along the lines recently developed for the three-dimensional 
case.\cite{Pi-S-98}

It is further clear that the criticisms recently raised to the fermionic 
T-matrix approximation when applied to the normal phase in three dimensions 
(which in the strong-coupling limit has been proved to reduce to the Born 
approximation as far as the scattering between composite bosons is concerned 
\cite{Pi-S-98}) are even more appropriate for the two-dimensional case where 
the Born approximation fails completely.
This remark makes it somewhat questionable the use of the fermionic 
(self-consistent) T-matrix approximation in two dimensions,\cite{KKK,EN} 
to describe the tendency toward the formation of preformed pairs above the 
superconducting critical temperature, at least as the strong-coupling (bosonic)limit is approached.

\acknowledgments

One of us (P.P.) gratefully acknowledges receipt of a postdoctoral research 
fellowship from the Italian INFM under contract PRA-HTCS/96-99.


\begin{references}
\bibitem{footnote-1} It can be also verified that the hydrodynamic relation 
$\mu_{B} = m_{B} s^{2}$ (where $s$ is the sound velocity) noted by Keldysh and 
Kozlov in Ref.~[5], is recovered from the analytic solution of Ref.~[6]
both in three and two dimensions.
\end{references}
\end{document}